\documentstyle[onecolumn,epsfig]{mn}
\newif\ifAMStwofonts

\usepackage[english]{babel}
\usepackage{epsfig}
\def\simlt{\mathrel{\rlap{\lower 3pt\hbox{$\sim$}}\raise 2.0pt\hbox{$<$}}}
\def\simgt{\mathrel{\rlap{\lower 3pt\hbox{$\sim$}} \raise 2.0pt\hbox{$>$}}}

\def\Msun{M_{\odot}}

\newcommand{\lsim}{\,\lower2truept\hbox{${<\atop\hbox{\raise4truept\hbox{$\sim$}}}$}\,}
\newcommand{\gsim}{\,\lower2truept\hbox{${>\atop\hbox{\raise4truept\hbox{$\sim$}}}$}\,}

\def\lsim{\,\lower2truept\hbox{${< \atop\hbox{\raise4truept\hbox{$\sim$}}}$}\,}
\def\gsim{\,\lower2truept\hbox{${> \atop\hbox{\raise4truept\hbox{$\sim$}}}$}\,}
\def\Ohat{{\widehat \Omega}}

\def\deg{\ifmmode^\circ \else$^\circ $\fi}    
\def\arcs{\ifmmode {'' }\else $'' $\fi}     
\def\arcm{\ifmmode {' }\else $' $\fi}     
\def\buildrel#1\over#2{\mathrel{\mathop{\null#2}\limits^{#1}}}
\def\mper{\ifmmode \buildrel m\over . \else $\buildrel m\over .$\fi}
\def\hper{\ifmmode \rlap.^{h}\else $\rlap{.}^h$\fi}
\def\sper{\ifmmode \rlap.^{s}\else $\rlap{.}^s$\fi}
\def\arcsper{\ifmmode \rlap.{' }\else $\rlap{.}' $\fi}
\def\arcmper{\ifmmode \rlap.{'' }\else $\rlap{.}'' $\fi}

\def\mincir{\ \raise -2.truept\hbox{\rlap{\hbox{$\sim$}}\raise5.truept	
\hbox{$<$}\ }}								%
\def\magcir{\ \raise -2.truept\hbox{\rlap{\hbox{$\sim$}}\raise5.truept	%
\hbox{$>$}\ }}								%

\title[Implications of the new generation of CMB spectrum space experiments]
{Mapping the thermal history of the Universe with the
new generation of CMB spectrum space experiments}

\author[C.~Burigana and R.~Salvaterra]
{C.~Burigana$^1$ and
R.~Salvaterra$^2$\\ 
$^1$IASF/CNR, Istituto di Astrofisica Spaziale e Fisica Cosmica,
Sezione di Bologna, \\
Consiglio Nazionale delle Ricerche, 
Via Gobetti 101, I-40129 Bologna, Italy \\
$^2$SISSA/ISAS, Astrophysics Sector, Via Beirut, 4, I-34014 Trieste, Italy}

\date{Submitted to MNRAS, 12 September 2003.}

\pagerange{\pageref{firstpage}--\pageref{lastpage}}
\pubyear{2003}

\begin{document}

\maketitle

\label{firstpage}
\footnotetext{E-mail: burigana@bo.iasf.cnr.it ; salvater@sissa.it}
\begin{abstract}
We have studied the implications 
of the new generation of CMB spectrum space experiments 
for our knowledge of the thermal
history of the Universe. 

The combination of two experiments with the sensitivity and the  
frequency coverage jointly forseen for {\it DIMES} and FIRAS~II will be able to 
greatly change our vision of the capability of the CMB spectrum 
information to constrain physical processes at different cosmic ages.
The limits on the energy dissipations at the all cosmic times
accessible to CMB spectrum investigations ($z \lsim z_{therm}$)
could be improved by about two order of magnitudes and even 
dissipation processes with $\Delta\epsilon/\epsilon_i \sim 10^{-6}$
could be detected and possibly accurately studied.

With a joint analysis of two dissipation processes occurring 
at different epochs, we demonstrated that 
the joint performances of {\it DIMES} and FIRAS~II would allow 
to accurately recover the two amounts of energy 
exchanged in the primeval plasma and to constrain quite well also the 
epochs of the two processes even when 
possible imprints from free-free distortions are taken into 
account. All the three distortion parameters could be accurately
reconstructed in this perspective: 
the sensitivity at 95 per cent CL 
is $\simeq (5-9) \times 10^{-7}$
for the two values of $\Delta\epsilon/\epsilon_i$ and of
$\simeq 10^{-7}$ for the free-free distortion parameter $y_B$. 
These results are possible because such levels of accuracy on a so wide 
frequency range allow to remove the approximate degeneracy both between
free-free and Bose-Einstein (BE) like distortions and between 
Comptonization and BE-like distortions that remain 
in presence of future significant improvements 
only at $\lambda \gsim 1$~cm or at $\lambda \lsim 1$~cm, respectively.  
The sensitivity on 
$\Delta\epsilon/\epsilon_i$, mainly determined by 
a FIRAS~II-like experiment, improves by a factor $\simeq 1.5$ 
by adding the information from a {\it DIMES}-like experiment,
while the sensitivity on 
$y_B$, mainly determined by 
a {\it DIMES}-like experiment, improves by a factor $\simeq 1.3-2.6$, 
by adding the information from a FIRAS~II-like experiment. 

Finally, we discussed the different signatures 
imprinted on the CMB spectrum by 
some late astrophysical and particle decay models 
recently proposed in the literature and possibly related
to the reionization of the Universe 
indicated by {\it WMAP}, and 
compared them with the sensitivity of such classes of CMB space 
spectrum experiments. 

\end{abstract}


\section{Introduction}

As widely discussed in many papers, the spectrum of the 
Cosmic Microwave Background (CMB) carries unique informations on physical
processes occurring during early cosmic epochs
(see e.g. Danese \& Burigana 1993 and references therein).
The comparison between models of CMB spectral distortions
and CMB absolute temperature measures can constrain the
physical parameters of the considered dissipation processes.
By jointly considering distortions generated in a wide range of 
early or intermediate cosmic epochs and at late cosmic epochs, 
we recently discussed 
the implications of the current CMB spectrum data 
(Salvaterra \& Burigana 2002)  as well as the great 
improvements in the knowledge of dissipation processes,
in particular on those possibly occurred 
at early and intermediate epochs, achievable
with a significant progress in the experiments at long wavelength, 
$\lambda \gsim 1$~cm, (Burigana \& Salvaterra 2003),
such that expected from a {\it DIMES}-like project (Kogut 1996, 2003).

Fixsen \& Mather (2002) recently proposed a new generation 
of millimeter and submillimeter spectrum experiment, FIRAS~II, aimed to 
improve by a factor $\sim 100$ the accuracy of the COBE/FIRAS  
measures.

While in the last decade many efforts have 
been~\footnote{http://lambda.gsfc.nasa.gov} 
(see e.g. Bennett et al. 1996, 2003 and references therein)
and are currently~\footnote{http://astro.estec.esa.nl/Planck/}  
(see e.g. Mandolesi et al. 1998, Puget et al. 1998, Tauber 2000
and references therein)
dedicated to greatly improve our knowledge 
of the whole sky CMB anisotropies since their COBE/DMR discovery 
(Smoot et al. 1992), the combination of high precision and large 
frequency coverage 
jointly forseen for {\it DIMES} and FIRAS~II opens a completely new 
perspective also in CMB spectrum studies, as we will discuss 
in this paper. 

In section~2
we briefly summarize the general properties of the
CMB spectral distortions and the main physical informations 
that can be derived from the comparison with the observations.
In section~3 we briefly report on the performances forseen for 
the {\it DIMES} and FIRAS~II experiments
and describe the generation of the simulated observations 
used in this work.
The implications of such experiments 
for our knowledge of the thermal history of the Universe
are presented in section~4:
models including an undistorted spectrum and distorted spectra 
in the case of a single kind of spectral distortion or by jointly 
considering three kinds of spectral distortions 
are compared with data simulated according to the 
sensitivities quoted for the above experiments
(the joint analysis of two spectral 
distortions is reported in Appendix~A for completeness).
Section~5 is dedicated to a comparison between some different 
classes of astrophysical and particle processes recently discussed in the 
literature occurring at relatively late epochs, able to distort the spectrum
at different wavelength regions, and, possibly, to
significantly contribute to the reionization of the Universe.
Finally, we discuss the results and draw our main conclusions in 
section~6.

\section{Theoretical framework}

The CMB spectrum emerges from the thermalization redshift, 
$z_{therm} \sim 10^6 - 10^7$, 
with a shape very close to a Planckian one, 
owing to the strict coupling between radiation and matter through
Compton scattering and photon production/absorption processes, 
radiative Compton and Bremsstrahlung,
which were extremely efficient at early times 
and able to re-establish a blackbody (BB) spectrum 
from a perturbed one
on timescales much shorter than the expansion time (see e.g. 
Danese \& De~Zotti 1977).
The value of $z_{therm}$ (Burigana, Danese \& De~Zotti 1991a)
depends on the baryon density (in units of the critical density),
$\Omega_b$, 
and the Hubble constant, $H_0$, through the product 
$\Ohat_b =\Omega_b (H_{0}/50)^2$ ($H_0$ expressed in Km/s/Mpc). 

On the other hand, physical processes occurring at redshifts $z < z_{therm}$ 
may lead imprints on the CMB spectrum.


The timescale for the achievement of
kinetic equilibrium between radiation and matter
(i.e. the relaxation time for the photon spectrum), $t_C$, is
\begin{equation}
t_C=t_{\gamma e} {m c^{2}\over {kT_e}} \simeq 4.5 \times 10^{28} 
\left( T_{0}/2.7\, K \right)^{-1} \phi^{-1} \Ohat_b^{-1}
\left(1+z \right)^{-4} \sec \, ,
\end{equation}
where $t_{\gamma e}= 1/(n_e \sigma _T c)$ is the photon--electron collision
time, $\phi = (T_e/T_r)$, $T_e$ being the electron temperature and
$T_r=T_{0}(1+z)$;
$kT_e/mc^2$ is the mean fractional change of photon energy in a scattering
of cool photons off hot electrons, i.e. $T_e \gg T_r$;
$T_0$ is the present radiation temperature related
to the present radiation energy density by $\epsilon _{r0}=aT_0^4$;
a primordial helium abundance of 25\% by mass is here assumed.

It is useful to introduce the dimensionless time variable $y_e(z)$ defined by
\begin{equation}
y_e(z) = \int^{t_0}_{t} {dt \over t_C}
=\int^{1+z}_{1} {d(1+z) \over 1+z} {t_{exp}\over t_C} \, ,
\end{equation}
where $t_0$ is the present time and
$t_{exp}$ is the expansion time given
by
%
\begin{equation}
t_{exp} \simeq   6.3\times 10^{19} \left({T_0 \over 2.7\, K}\right)^{-2}
(1+z)^{-3/2} \left[\kappa (1+z) + (1+z_{eq})
-\left({\Omega_ {nr} -1 \over \Omega_ {nr}}\right)
\left({1+z_{eq} \over 1+z}\right) \right]^{-1/2}
\sec \, ,
\end{equation}
$z_{eq} = 1.0\times 10^4 (T_{0}/2.7\, K)^{-4}\Ohat _{nr}$
being the redshift of
equal non relativistic matter and photon energy densities
($\Omega _{nr}$ is the density of non relativistic matter in units of critical
density); $\kappa = 1 + N_\nu (7/8)
(4/11)^{4/3}$, $N_\nu$ being the number of relativistic, 2--component,
neutrino species (for 3 species of massless neutrinos, $\kappa \simeq 1.68$),
takes into account the
contribution of relativistic neutrinos to the dynamics of the
Universe\footnote{Strictly speaking the present ratio of neutrino to
photon energy densities, and hence the value of $\kappa$, is itself a
function of the amount of energy dissipated. The effect, however,
is never very important and is negligible
for very small distortions.}.

Burigana, De~Zotti \& Danese 1991b have reported on
numerical solutions of the Kompaneets equation (Kompaneets 1956)
for a wide range of values of the relevant parameters
%
%
and accurate analytical representations of these numerical solutions,
suggested in part by the general properties of the Kompaneets 
equation and by its well known asymptotic solutions,
have been found (Burigana, De~Zotti \& Danese 1995).

The CMB distorted spectra depend on at least
three main parameters: the fractional amount of energy exchanged between
matter and radiation, $\Delta\epsilon / \epsilon_i$,
$\epsilon _i$ being the radiation energy density before the energy injection,
the redshift $z_h$ at which the heating occurs, and the
baryon density $\Ohat_b$.
The photon occupation number can be then expressed in the form
\begin{equation}
\eta = \eta (x; \Delta\epsilon / \epsilon_i, y_h, \Ohat_b) \, ,
\end{equation}
where $x$ is the dimensionless frequency $x = h\nu/kT_{0}$
($\nu$ being the present frequency),
and $y_h \equiv y_e(z_h)$ characterizes the epoch when the energy dissipation
occurred, $z_h$ being the corresponding redshift
(we will refer to $y_h \equiv y_e(z_h)$ computed assuming $\phi=1$, so that the epoch 
considered for the energy dissipation does not depend on the 
amount of released energy).
The continuous behaviour of the distorted spectral shape with $y_h$ can be
in principle used also to search for constraints on the 
epoch of the energy exchange.
Of course, by combining the approximations describing the distorted spectrum
at early (Sunyaev \& Zeldovich 1970)
and intermediate epochs with the Comptonization distortion 
expression (Zeldovich \& Sunyaev 1969; Zeldovich, Illarionov \& Sunyaev 1972)
describing late distortions, it is possible to jointly treat
two heating processes (see Burigana et al. 1995 and Salvaterra \&
Burigana 2002 and references therein for a more exhaustive discussion).

In this work
simulated  measures of the CMB absolute temperature 
are compared with the above models
of distorted spectra
by using a standard $\chi^2$ analysis. 

We determine the limits on the amount of energy possibly injected 
in the cosmic background at arbitrary primordial epochs corresponding to a
redshift $z_h$ (or equivalently to $y_h$).
This topic has been discussed in several works
(see e.g. Burigana et al. 1991b,
Nordberg \& Smoot 1998, Salvaterra \& Burigana 2002). 
We apply here the method of comparison with the theoretical models 
described in Salvaterra \& Burigana 2002 
and Burigana \& Salvaterra 2003,
by investigating the possibility of properly combining
data simulated according to the {\it DIMES} and FIRAS~II performances 
respectively at long and short wavelengths. 

We will consider the recent improvement in the
calibration of the FIRAS data, that sets the CMB scale temperature at
$2.725\pm0.002$~K at 95 per cent confidence level (CL) (Mather et al. 1999). 

Then, we study the combined effect of two different heating processes
that may have distorted the CMB spectrum at different epochs.
This hypothesis has been also taken into account by 
Nordberg \& Smoot 1998, who fit
the observed data with a spectrum distorted by a single heating at $y_h=5$,
a second one at $y_h\ll 1$ and by free-free emission, obtaining limits on the
parameters that describe these processes.
As in Salvaterra \& Burigana 2002, we extend their analysis by
considering the full range of epochs for the early and 
intermediate energy injection
process, by taking advantage of the analytical representation
of spectral distortions at intermediate redshifts (Burigana et~al. 1995).

Long wavelength measures are particularly sensitive to the 
free-free distortion due to its well known dependence 
on the frequency. At moderately long wavelengths, where 
quite accurate observations can be carried out,   
$(T_{th}-T_0\phi_i)/T_0 \simeq y_B/x^2 \, $.
Here $\phi_i$ is the ratio between the electron and radiation 
temperature before the beginning of the dissipation process,
$y_B$ is the so called free-free distortion 
parameter (Burigana et al. 1995), and $T_{th}$ is the (frequency dependent)
equivalent thermodynamic temperature.
At very long wavelengths the equivalent thermodynamic temperature
approaches the electron temperature because of the 
very high bremsstrahlung efficiency.

The relationship between free-free distortion and Comptonization
distortion is highly model dependent,
being related to the details of the thermal history at late
epochs (see e.g. Danese \& Burigana 1993, Burigana et al. 1995),
and can not be simply represented by integral parameters.

In the combined analysis of the implications of future CMB spectrum 
experiments for energy dissipations at different cosmic times and
free-free distortions we avoid to physically link 
Comptonization and free-free distortions. Eq.~(4) generalizes then to 
\begin{equation}
\eta = \eta (x; \Delta\epsilon / \epsilon_i (y_h), y_h, 
\Delta\epsilon / \epsilon_i (y_h \ll 1), y_B, \Ohat_b) \, .
\end{equation}

Comptonization and free-free emission could be also produced 
in several astrophysical scenarios. For example, 
ionized halos in the early stages of galaxy formation (Oh 1999)
can produce a relevant free-free distortions 
and coupled Comptonization and free-free distortions 
can be produced in the reionization epoch, indicated by
the {\it WMAP} data (Kogut et al. 1993). 


It is also possible to extend the limits 
on $\Delta\epsilon/\epsilon_i$ for heatings
occurred at $z_h>z_1$, where $z_1$ is the redshift corresponding
to $y_h = 5$, when the Compton
scattering was able to restore the kinetic equilibrium between matter and
radiation on timescales much shorter than the expansion time
and the evolution on the CMB spectrum can be easily studied
by replacing the full Kompaneets equation with the differential
equations for the evolution of the electron temperature 
and the chemical potential.
This study can be performed by using the
simple analytical expressions by Burigana et al. 1991b
instead of numerical solutions.


To evaluate the scientific impact represented by 
the future experiment improvements, we
create different data sets 
simulating the observation of a distorted and non distorted spectra  
from a space experiments like {\it DIMES} and FIRAS~II 
through the method described in section~3.1.

Each data set will be then compared to models of distorted spectra by using 
the method described in Salvaterra \& Burigana 2002
(see also Burigana \& Salvaterra 2000 for the details of the code)
to recover the values of the parameters appearing in Eq.~(5).


For simplicity, we restrict 
to the case of a baryon density
$\Ohat_b = 0.05$ our analysis 
of the implications for the thermal history of the Universe,
but the method can be simply applied to different 
values of $\Ohat_b$.

\section{Future experiments}

The CMB spectrum experiments considered here are 
dedicated to improve our knowledge both at the same and at
longer wavelengths with respect to the FIRAS frequency coverage 
($1$~cm~$\gsim \lambda \gsim 0.01$~cm).

As representative cases, and without the ambition to cover the whole 
set of planned experiments, we briefly 
refer here to the {\it DIMES} experiment from space (Kogut 1996)
designed to reach an accuracy close to that of FIRAS 
up to $\lambda \simeq 15$~cm and to the FIRAS~II experiment (Fixsen \& 
Mather 2002) which will allow a sensitivity improvement by a factor $\sim 
100$ with respect to FIRAS. 

{\it DIMES}~\footnote{http://map.gsfc.nasa.gov/DIMES/index.html} (Diffuse 
Microwave Emission Survey) is a space mission submitted to the NASA in
1995, designed to measure very accurately the CMB spectrum at wavelengths
in the range $\simeq 0.5 - 15$~cm (Kogut 1996).

{\it DIMES} will compare the spectrum of each 10 degree pixel on the sky to a
precisely known blackbody to precision of $\sim 0.1$~mK,
close to that of FIRAS ($\simeq 0.02 - 0.2$~mK). 
The set of receivers is given
from cryogenic radiometers 
with instrument emission cooled to 2.7~K
operating at six frequency bands 
about 2, 4, 6, 10, 30 and 90~GHz using a single external blackbody 
calibration target common to all channels to minimize the calibration
uncertainty. 
The {\it DIMES} design is driven by the
need to reduce or eliminate systematic errors from instrumental artifacts.
The {\it DIMES} sensitivity represents an
improvement by a factor better than 300 with respect to 
previous measurements at cm wavelengths.

Fixsen \& Mather (2002) described the fundamental guidelines
to significantly improve CMB spectrum measures at $\lambda \lsim 1$~cm.
A great reduction of the residual noise of cosmic rays, dominating
the noise of the FIRAS instrument, can be obtained by eliminating
the data on-board co-add process or applying deglitching before co-adding,
by reducing the size of the detectors and by using 
``spiderweb'' bolometers or antenna-coupled microbolometers.
They are expected to show a very low noise when cooled below 1~K
while RuO  sensors can reduce to 0.1~mK the read noise of thermometers.
The Lagrangian point L2 of the Earth-Sun system is, of course, the
favourite ``site'' for FIRAS~II. Also, the calibration can be  
improved by order or magnitudes 
with respect to that FIRAS by reducing the contribution to
the calibrator  reflectance  of light from the diffraction
at the junction between the calibrator and the horn.
A complete symmetrical construction of the instrument is 
recommended. This allows the cross-check between calibrators 
and between calibrators and the sky and to realize 
``an end-to-end calibration and performance test before lunch''.
According to the authors, FIRAS~II can be designed to have 
a frequency coverage from 60 to 3600~GHz  
(i.e. from 5~mm to 83~$\mu$m) with a spectral 
resolution $\nu / \Delta \nu < 200$ and sensitivity 
in each channel about 100 times better than FIRAS.  

\subsection{Generation of simulated data sets} 

We collect different data sets, 
simulating measurements of different CMB spectra, distorted or not, at the 
frequency ranges of the considered experiments. 

For the cases of distorted spectra
we calculate the theoretical temperature of the CMB spectrum
as discussed in the previous section.
Of course, 
the equivalent thermodynamic temperature
held obviously constant at all the frequencies
for the case of a non distorted spectrum. 

The theoretical temperatures
are then fouled to simulate real measurements affected by instrumental noise.
The simulated temperature $T_{obs}$ at the frequency $\nu$ is
given by $T_{obs}(\nu)=T_{teor}(\nu)+n(\nu)\times \mbox{err}(\nu) \, ,$
where $T_{teor}(\nu)$ is the theoretical temperature at the frequency $\nu$
and err($\nu$) is the expected rms error (at 1~$\sigma$) 
of the experiment at this frequency.
The numbers $n$ are a set of random numbers generated according to a Gaussian
distribution with null mean value and unit variance with the routine
GASDEV by Press et al. 1992 (\S 7).

The considered frequency channels are those of {\it DIMES} 
and those achievable with the spectral resolution forseen
for FIRAS~II.
For sake of simplicity and 
differently from the analysis carried out in Salvaterra \& Burigana 2002,
we do not consider here the wavelength region 
at $\lambda \le 800 \mu$m, accessible to FIRAS~II,
and neglect the problem of separating the signatures of CMB 
spectral distortions, dominant at $\lambda \gsim 1$~mm, from the contribution 
from Far-IR galaxies, dominant at $\lambda \lsim 500 \mu$m.

\subsection{Simulated data sets}\label{dati_dimes}

We generate a set (S-BB) of simulated data in the case of a blackbody
spectrum at a temperature of 2.725~K 
in order to evaluate the joint capability of
two experiments like {\it DIMES} and FIRAS~II 
to improve the constraints on the 
amount of the energy injected in the cosmic radiation field.
The analysis of this case is in fact directly 
comparable with the results obtained from the fit to the
currently available data.

Of course, it is extremely important to evaluate the capability of 
so accurate experiments to detect and, possibly, precisely measure 
the spectral distortion parameters, the fractional  
energy exchanges and the dissipation epochs.
To this aim, we generate other three sets of simulated data
(respectively, S-E5C, S-E1.5C, S-E0.1C)
assuming an instantaneous ($\delta y_h  \ll y_h$) 
earlier energy dissipation with
$\Delta \epsilon /\epsilon _i (y_h) = 3\times10^{-6}$ 
($\simeq \mu/1.4$, where $\mu$ is the chemical potential 
of the Bose-Einstein (BE) formula
describing the high frequency tail of the spectrum 
for distortions at sufficiently high redshifts)
occurring at $y_h = 5$ or at $y_h = 1.5$ 
or at $y_h = 0.1$ 
plus a later Comptonization distortion
associated to an energy dissipation
$\Delta \epsilon /\epsilon _i (y_h \ll 1) = 3\times10^{-6} \simeq 4 u$
(here $u$ is the ``energy'' Comptonization parameter).
For simplicity, we assume here $y_B = 0$ (see Sect.~5 for 
a discussion of combined Comptonization and free-free distortions
on the basis of some specific models).   
All these distorted spectra are computed by setting $\Omega_b=0.05$ 
and $H_0=50$~Km/s/Mpc.

These above simulated data set will be compared with models
including spectral distortions generated by a 
dissipation process occurred at very different cosmic 
epochs, on a grid represented 
by the dimensionless time $y_h =$ 
 5, 4, 3, 2, 1, 0.5, 0.25, 0.1, 0.05, 0.025, 0.01, and 
$y_h\ll1$, plus/or a Comptonization distortion, plus/or a free-free 
distortion, as described in Sect.~2, in order to analyse 
the quality of the recovery of the input parameters 
achievable with the considered experiments.

\section{Results}\label{sec:results}

\subsection{Fits to simulated data: non distorted spectrum}

\subsubsection{Analysis in terms of a single spectral distortion}

We fit the simulated data set S-BB with a spectrum distorted by a
single energy injection at different values of $y_h$ 
or by a free-free distortion in order 
to recover the value of $\Delta\epsilon/\epsilon_i$
or of $y_B$, expected to be null, and the limits (at 95 per cent of CL)  
on them.
The results are reported in Table~1 for a representative
set of the considered dissipation epochs.
The table shows also the results 
obtained by using the FIRAS data alone  
combined with the long wavelength data 
accumulated in the last two decades 
(see also Salvaterra \& Burigana 2002), 
the results achievable
by combining the FIRAS data with the subset of the
data set S-BB achievable by {\it DIMES} alone
(see also Burigana \& Salvaterra 2003), and,
finally, the results achievable
by exploiting the subset of 
data set S-BB achievable by FIRAS~II alone.

The sensitivities reported in Table~1
for a given distortion parameter
apply in the case in which quite strong priors 
can be assumed on the other two distortion parameters.
In any case,
FIRAS~II alone
or {\it DIMES} alone either is able 
to improve by more than one order of magnitude 
the current limits on the free-free distortions 
and on the energy exchanges.
It is evident from the table that in this case 
FIRAS~II alone is able to provide the best constraints on 
the energy exchanges 
(with a sensitivity improvement
by a factor $\sim 100$)
while {\it DIMES} alone 
is able to provide the best constraints on the
free-free distortions
(with a sensitivity improvement
by a factor $\simeq 500$), without 
any significant improvement from the
combination of the two experiments.

\begin{table}
\begin{center}
\begin{tabular}{lccccc}
\hline
\hline
\\
Data set & \multicolumn{4}{c}{$(\Delta\epsilon/\epsilon_i)/10^{-7}$} & $y_B/10^{-7}$ \\
\cline{2-5}
\\
 & heating at & heating at & heating at &  heating at &free-free dist. at \\
 & $y_h=5$ & $y_h=1.5$ & $y_h=0.5$ & $y_h\ll 1$ & $y_h\ll 1$ \\
\hline
\\ 
F & 20$\pm$533 & 25$\pm$453 & 25$\pm$359 & 28$\pm$233 & 272$\pm$920 \\
F + R & 64$\pm$532 & 39$\pm$453 & 28$\pm$359 & 28$\pm$233 & 
-455$\pm$424 \\
\\
\hline
\\
D + F & -1.92$\pm$18.12 & -4.5$\pm$38.1 & -3.2$\pm$106.9 & 
3.08$\pm$210 & -0.02$\pm$0.88 \\
\\
\hline 
\\
F-II & 1.28$\pm$3.95 & 1.06$\pm$3.35  & 0.83$\pm$2.66 & 0.60$\pm$2.08 
& -3.26$\pm$7.33 \\
D + F-II & 1.13$\pm$3.85 & 1.02$\pm$3.34 & 0.82$\pm$2.66 & 
0.60$\pm$2.08 & -0.07$\pm$0.87 \\
\\
\hline
\end{tabular}
\end{center}
\caption{Fits to a single spectral distortion by considering different sets of
real and/or simulated data. 
F: FIRAS; R: recent measures at $\lambda \gsim 1$~cm; D: {\it DIMES}; 
F-II: FIRAS II.
An underlying blackbody spectrum is here assumed
for the simulated data sets.}
\label{tab:1}
\end{table}

\subsubsection{Joint analysis of three kinds of spectral distortions}

We discuss here the possibility to significantly improve
the constraints on the spectral distortion parameters 
in the more general case of
a joint analysis of an early/intermediate dissipation process
occurred at a given $y_h$,  
of a late dissipation process ($y_h \ll 1$), and of 
a free-free distortion.
We exploit the same data sets of the previous subsection
but without any prior on the distortion parameters.

The results are reported in Table~2 (95 per cent CL)
for a representative
set of the earlier dissipation epoch $y_h$
and displayed in Fig.~1 for the whole range of $y_h$.

Clearly, the sensitivities on each distortion parameter
are worse than the ``analogous'' one in Table~1, but 
the relative improvements with respect to the
current observational status are similar.

Although FIRAS~II ({\it DIMES}) still 
plays the most relevant role for 
the energy exchanges (for the free-free distortions),
it is interesting to note that in this case
the spectral distortion parameter recovery 
appreciably improves by combining the two experiments.
This is because such levels of accuracy on a so wide 
frequency range allow to remove the approximate degeneracy both between
free-free and BE-like distortions and between 
Comptonization and BE-like distortions that remain 
in presence of future significant improvements 
only at $\lambda \gsim 1$~cm or at $\lambda \lsim 1$~cm, respectively.  

This point will be discussed again in the next subsections,
where the analysis of the 
capability of such experiments 
to measure spectral distortions possibly
present in the CMB spectrum and the corresponding dissipation 
epochs is presented.

\begin{table}
\begin{center}
\begin{tabular}{lccccc}
\hline
\hline
\\
Data set & \multicolumn{4}{c}{$(\Delta\epsilon/\epsilon_i)/10^{-7}$} & $y_B/10^{-7}$ \\
\cline{2-5}
\\
 & heating at & heating at & heating at &  heating at &free-free dist. at \\
 & $y_h=5$ & $y_h=1.5$ & $y_h=0.5$ & $y_h\ll 1$ & $y_h\ll 1$ \\
\hline
\\
F & 414$\pm$1561 & & & -85$\pm$546 & 635$\pm$1531 \\
F & & 353$\pm$1334 & & -88$\pm$560 & 614$\pm$1473 \\
F & & & 286$\pm$1091 & -99$\pm$601 & 583$\pm$1392 \\
F + R & -465$\pm$1056 & & & 163$\pm$436 & -540$\pm$469 \\
F + R & & -418$\pm$922 & & 176$\pm$452 & -542$\pm$459 \\
F + R & & & -343$\pm$789 & 189$\pm$492 & -521$\pm$453 \\
\\
\hline
\\
D + F & -25.6$\pm$56.02 & & & 68.51$\pm$231 & -1.12$\pm$2.64 \\
D + F & & -35.0$\pm$89.1 & &71.6$\pm$240 & -0.67$\pm$1.94 \\
D + F & & & -33.6$\pm$176.4 & 63.5$\pm$275 & -0.14$\pm$1.18 \\
\\
\hline
\\
F-II & -3.43$\pm$13.87 & & & 1.82$\pm$6.20 & -5.22$\pm$12.06 \\
F-II & & -3.05$\pm$12.22 & & 1.92$\pm$6.56 & -5.14$\pm$11.77 \\
F-II & & & -2.64$\pm$10.28 & 2.13$\pm$7.15 & -5.03$\pm$11.24 \\
D + F-II & 0.51$\pm$9.07 & & & 0.36$\pm$4.72 & -0.38$\pm$1.00 \\
D + F-II & & 0.71$\pm$8.00 & &0.20$\pm$4.94 & -0.44$\pm$0.90 \\
D + F-II & & &0.68$\pm$6.85 & 0.10$\pm$5.42 & -0.54$\pm$0.88 \\
\\
\hline
\end{tabular}
\end{center}
\caption{Fits to three kinds of spectral distortions, jontly considered, for 
different real and/or simulated data.
F: FIRAS; R: recent measures at $\lambda \gsim 1$~cm; D: {\it DIMES}; 
F-II: FIRAS II.
An underlying blackbody spectrum is here assumed
for the simulated data sets.}
\label{tab:3}
\end{table}

\begin{figure*}
\epsfig{figure=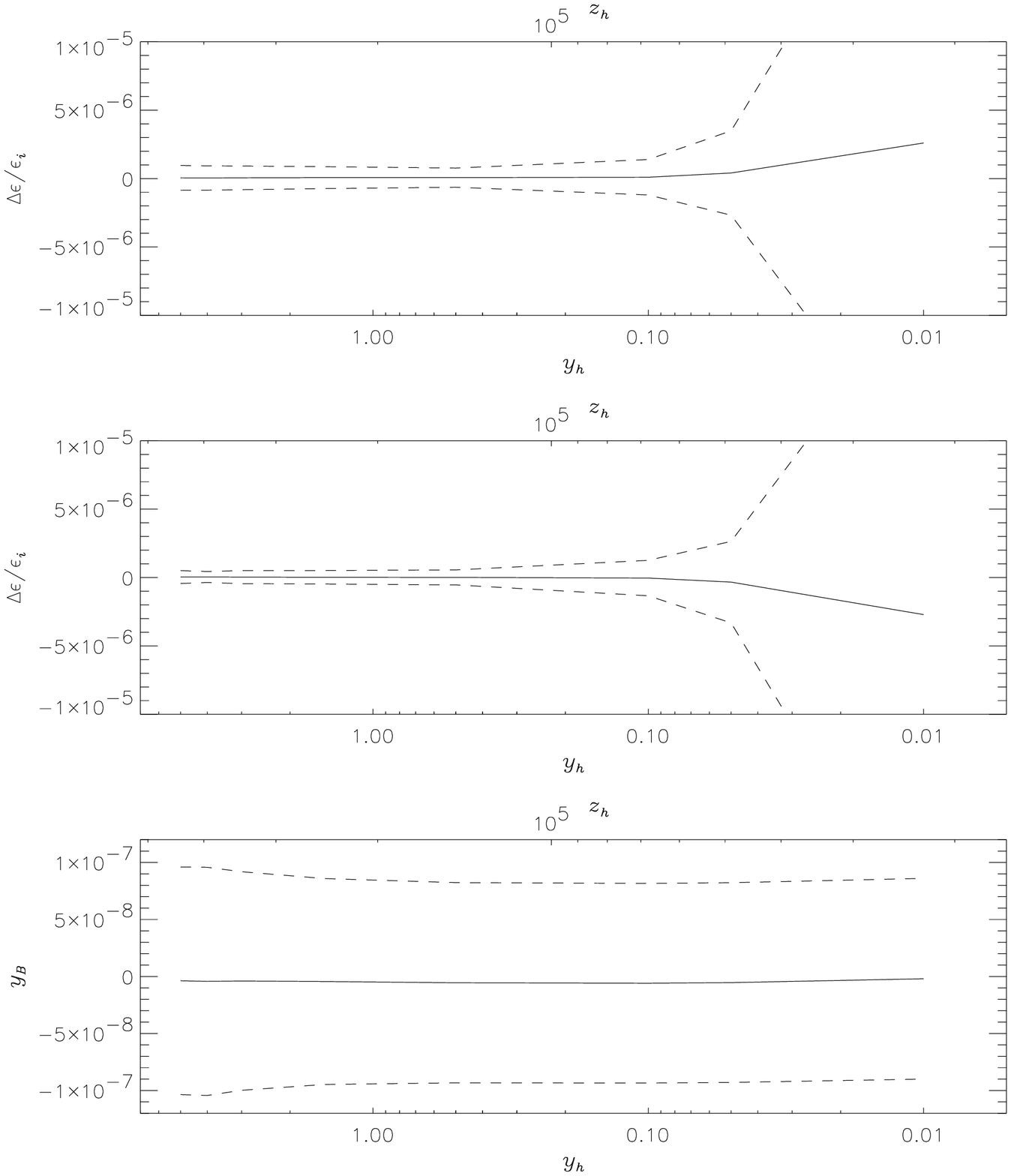,height=12cm,width=17cm}
\caption{Constraints at 95 per cent CL 
on two energy exchanges at different epochs and on the free-free distortion,
jointly considered, as could be
in principle derived by combining a {\it DIMES}-like experiment
with a FIRAS~II-like experiment. 
An underlying blackbody spectrum is here assumed.
Solid lines: best fit; dashes: upper and lower limits.
Top panel: constraints on 
$\Delta\epsilon/\epsilon_i (y_h)$ allowing for 
a later energy exchange at $y_h \ll 1$ and for a 
free-free distortion.
Middle panel: constraints on 
$\Delta\epsilon/\epsilon_i (y_h \ll 1)$ allowing for 
a previous energy exchange occurring at $y_h$ and for a 
free-free distortion.
Bottom panel: constraints on 
$y_B$ allowing for a first energy exchange occurring at $y_h$ 
and a second energy exchange at $y_h \ll 1$.
See also the text.}
\end{figure*}

\subsection{Fits to simulated data: distorted spectra}

In this subsection we exploit the  
simulated data S-E5C, S-E1.5C, and S-E0.1C (see Sect.~3.2)
to evaluate the quality of spectral distortion parameter
recovery achievable with the considered experiments.

\subsubsection{Combining {\it DIMES} and FIRAS~II}

We first consider the information
achievable by combining {\it DIMES} and FIRAS~II.

In the case of the data set S-E5C, 
a fit in terms of two energy exchanges
at the epochs assumed in the input model
($y_h = 5$ and $y_h \ll 1$)
shows that the values of 
$\Delta \epsilon /\epsilon _i$ can be accurately 
recovered: for example, in the considered test, we find 
$\Delta \epsilon /\epsilon _i (y_h =5) = (2.79 \pm 0.79 ) \times 10^{-6}$ 
and 
$\Delta \epsilon /\epsilon _i (y_h \ll 1) = (3.03 \pm 0.42 ) \times 10^{-6}$
(errors at 95 per cent CL)
with a $\chi^2$/d.o.f.~$=1.100$.
This results are not significantly affected by 
allowing also for a free-free distortion. We find in this case:
$\Delta \epsilon /\epsilon _i (y_h =5) = (2.71 \pm 0.90 ) \times 10^{-6}$, 
$\Delta \epsilon /\epsilon _i (y_h \ll 1) = (3.06 \pm 0.47 ) \times 10^{-6}$,
and the recovered 
$y_B = (-1.69 \pm 9.96) \times 10^{-8}$ is clearly 
in agreement with the null input value
($\chi^2$/d.o.f.~$=1.119$).

We verified also that a (wrong) interpretation of these simulated
data in terms of a single kind of spectral distortion
is clearly ruled out
(and the recovery of the spectral distortions parameters
is, of course, wrong).  
Even in the most favourite case of
a Comptonization distortion we find  
a $\chi^2$/d.o.f.~$\simeq 2$ (the $\chi^2$ goes
from $\simeq 58$ of previous fits to $\simeq 108$).
Analogously, an interpretation in terms of a single energy dissipation
plus a free-free distortion is ruled out
(we find $\chi^2$/d.o.f.~$=1.774$ in the most favourite case of
a Comptonization distortion coupled to a free-free distortion).

We then tried to estimate the epoch of the earlier 
energy injection. By assuming $y_h=2$ we find
a $\chi^2 \simeq 58$ but  
a degraded spectral distortion parameter recovery. 
In particular, the 
$\Delta \epsilon /\epsilon _i$ value
of the later process is very weakly affected, while 
that of the earlier process is underestimated of $\simeq 10$~\%
and a (wrong) small negative free-free detection,
$y_B = (-7.34 \pm 9.21) \times 10^{-8}$, appears, although 
clearly compatible with the input value $y_B = 0$ 
at $\sim 1.6 \sigma$. 

If we try a fit assuming $y_h =1$ the 
results, substantially unchanged in terms of $\chi^2$,
degrades further from the physical point of view: 
we find 
$y_B = (-1.14 \pm 0.89) \times 10^{-7}$, i.e. 
a sign no longer compatible (at $\simeq 2.6 \sigma$) 
with that of the, again only slightly affected,
recovered Comptonization distortion. 

To verify the physical trouble of such possible
sign differences we computed the
(negative) Comptonization distortion parameter 
produced by 
simple thermal histories (namely, with $\phi \simeq {\rm constant}$
and fully ionized medium) at pre-recombination epochs
or after the recombination 
able to generate free-free distortions in the range on  
the negative values found above.
For example, we find 
$u \lsim - 1.8 \times 10^{-5}$
by assuming a process in the redshift range $10^4-10^3$
and   
$u \lsim - 5.6 \times 10^{-7}$ for a process 
in the redshift range $10^3-10^2$, while
a process
in the redshift range $30-0$ can produce
such negative free-free distortions only
for unphysical matter cooling processes involving 
$\phi \lsim 0.2$. 
Since also a cooling process with significant ionization 
at $z \sim 10^3-10^2$ seems to be excluded by {\it WMAP},
the only possible scenarios producing different signs of 
$y_B$ and $u$ should involve or a pre-recombination
cooling process able to produce negative Comptonization 
and free-free distortions followed by a late heating
process with a very small free-free distortion and 
a significant Comptonization distortion able to 
properly compensate the previous one, or, in general, 
particularly ad hoc thermal histories. 
In general, these possible scenarios should involve 
accurate fine tunings and appear difficult and untenable
and should be discarded in favour of an early energy injection
at relevant $y_h$ followed by a late dissipation mechanism.

In spite of the insignificant variations of the 
$\chi^2$, the disagreement between the signs of free-free 
and Comptonization distortions due to the wrong assumption 
on the earlier process epoch, tends to 
``physically'' support the (right) interpretation 
of the S-E5C data set involving an energy dissipation
at $y_h$ certainly larger than 1 and, probably, larger than 2 or 3.
We then conclude that 
reliable estimates of the epochs of the possible dissipation processes 
are achievable by combining {\it DIMES}-like and 
FIRAS~II-like experiments, even for very small 
energy exchanges.

By exploiting the data set S-E1.5C, we find 
that a fit in terms of two energy exchanges
at the epochs assumed in the input model
($y_h = 1.5$ and $y_h \ll 1$)
allows to recover the values of 
$\Delta \epsilon /\epsilon _i$ 
with a sensitivity similar to that
quoted above for the data set S-E5C.
Again, allowing for a free-free distortion
does not represent a problem and models involving
a single kind of spectral distortion
or a single energy dissipation
plus a free-free distortion can be easily 
ruled out by simple $\chi^2$ arguments.
Concerning the determination of 
the epoch of the earlier energy injection,
in this case 
the comparison between the recovered signs of
Comptonization and free-free distortions
does not play a relevant role and  
the direct analysis of the
$\chi^2$ rules out values of $y_h$ smaller than $\simeq 0.1$.
 
Finally, we analysed the data set  
S-E0.1C. We note that the differences between a spectrum 
resulting from a dissipation at $y_h \simeq 0.1$
and that produced by a late ($y_h \ll 1$) process 
are significantly less evident that those
between a spectrum
resulting from a dissipation at relevant $y_h$ ($y_h \gsim 1$)
and that produced by a late ($y_h \ll 1$) process.
Consequently, we may expect a degradation of the
fit result quality. In fact, we find an appreciable
underestimation of the earlier energy dissipation and
an analogous overestimation 
of the later energy dissipation, and error bars significantly
larger that in the previous cases:
$\Delta \epsilon /\epsilon _i (y_h =0.1) = (2.31 \pm 1.34 ) \times 10^{-6}$ 
and 
$\Delta \epsilon /\epsilon _i (y_h \ll 1) = (3.59 \pm 1.33 ) \times 10^{-6}$
(errors at 95 per cent CL).
Note that, even in this unfavourite case, 
the recovered values of $\Delta \epsilon /\epsilon _i$
are in agreement with the input ones at $\simeq 1\sigma$ level.
Also in this case, allowing for a free-free distortion
does not degrade further the fit energy exchange recovery
and models involving
a single kind of spectral distortion
or a single energy dissipation
plus a free-free distortion can be directly 
ruled out by simple $\chi^2$ arguments.
A firm estimate of the epoch of the earlier energy injection
can not be achieved in this case:
fits involving models with an earlier energy injection
at $y_h \gsim 1$ do not show a significant increase
of the $\chi^2$. On the other hand, from the analysis
of the previous cases we know that the 
epoch of a possible energy injection at relevant $y_h$
can be quite well estimated from such high quality data.
Therefore, in the situation in which a fit to the data in 
terms of two energy injections at different epochs 
(plus, possibly a free-free distortion) does not allow
to estimate the epoch of the earlier process,
it is quite reasonable to argue that the earlier process
has been occurred at an epoch 
not particularly early (by looking at the results 
found for the data set S-E1.5C, we argue in this case 
an upper limit $y_h \lsim 1$).
  
We conclude that the combined information
contained in future {\it DIMES} and FIRAS~II experiments
will allow to clearly identify 
the presence of spectral distortions also in the
quite general case of a thermal histories involving
early/intermediate and late processes,  
to measure the corresponding parameters 
with a good precision even for small amounts
($\Delta \epsilon /\epsilon _i \sim 10^{-6}$) 
of dissipated energy, and to quite well constrain
the epochs at which the processes have been possibly occurred.
 
\subsubsection{FIRAS~II alone}

In the previous subsection we have considered the information
achievable by combining {\it DIMES} and FIRAS~II.
Burigana \& Salvaterra 2003 discussed the scientific
capabilities of accurate long wavelength measures coupled
to the available FIRAS data.
We then report here the results obtained by repeating the same 
analyses carried out above, but by considering the subsets of the data sets
S-E5C, S-E1.5C, and S-E0.1C achievable by using FIRAS~II alone.

In the case of the data set S-E5C,
we find that the two values of 
$\Delta \epsilon /\epsilon _i$ can be recovered with an accuracy only
slightly degraded with respect to that of the previous section
only by performing the 
fit in terms of two energy exchanges
at the epochs assumed in the input model 
(the resulting $\chi^2$/d.o.f.~$=1.136$ in this case)
while allowing for a free-free distortion implies a degradation
of a factor $\simeq 1.5$ in the $\Delta \epsilon /\epsilon _i$ 
recovery sensitivity. 
Models with a single spectral distortions
can be again ruled out; the same holds for 
models involving a single energy dissipation
plus a free-free distortion,
although with a $\chi^2$ increase less significant than 
that found in the previous subsection
(for example, for a fit in terms of a Comptonization distortion coupled to 
a free-free distortion  
we find $\chi^2$/d.o.f.~$=1.399$, to be compared with
the value of 1.774 found in the previous subsection).  

Results similar to those described above are found
by considering the data set S-E1.5C.
The main difference concerns the impact of free-free distortion:
by allowing for it in the fit we find in the test 
a significant underestimation ($\simeq 36$~\%, no longer compatible
at $1\sigma$ with the input one)
on the recovery of the earlier energy injection
($\Delta \epsilon /\epsilon _i (y_h =1.5) = (1.92 \pm 1.35 ) \times 10^{-6}$) 
and a corresponding (false) detection of a negative (at $\simeq 1.3 \sigma$)
free-free distortion
($y_B = (-0.823 \pm 1.248 ) \times 10^{-6}$)  
[while the effect on Comptonization
distortions is small; we find
$\Delta \epsilon /\epsilon _i (y_h \ll 1) = (3.37 \pm 0.72 ) \times 10^{-6}$,
errors at 95 per cent CL].
Clearly, by using only high frequency data, the 
two errors tends to compensate each other, because of their
signatures in the final spectrum.

The exploitation of the data set S-E0.1C
shows analogous results.
The above problem related to the impact of free-free distortion
is further enhanced and the parameter recovery sensitivity
is significantly degrades: we find
$\Delta \epsilon /\epsilon _i (y_h = 0.1) = (1.58 \pm 2.28 ) \times 10^{-6}$,
$y_B = (-0.517 \pm 1.199 ) \times 10^{-6}$,
$\Delta \epsilon /\epsilon _i (y_h \ll 1) = (4.18 \pm 2.20 ) \times 10^{-6}$
(errors at 95 per cent CL).
In addition, while models with a single spectral distortions
can be again directly ruled out, 
and the same holds for models without a late energy injection,
models involving a single late ($y_h \ll 1$) energy dissipation
plus a free-free distortion
can not be longer ruled out on the basis on a direct 
$\chi^2$ analysis but on the requirement
of having Comptonization and free-free distortions
with the same sign
(we recover in this case the values
$\Delta \epsilon /\epsilon _i (y_h \ll 1) \simeq (5.78 \pm 0.22) \times 10^{-6}$,
i.e. about corresponding to the sum the energy dissipated by the 
two processes in order to fit the higher frequency tail of the spectrum,
and a negative $y_B$ ($\simeq (-1.22 \pm 0.76 ) \times 10^{-6}$)
that tends to mimic the effect of an earlier energy dissipation).

Finally, we find that for all the three data sets only a poor
information on the epoch of the earlier energy dissipation 
can be achieved.

\subsubsection{FIRAS~II alone versus {\it DIMES} and FIRAS~II}

From the comparison between the results 
reported in the two previous subsections
we can draw the following conclusions.

A FIRAS~II-like experiment shows a sensitivity 
to the values of the two energy exchanges at different cosmic
times similar to that achieved by combining   
a FIRAS~II-like experiment and a {\it DIMES}-like experiment 
only in presence of a strong prior on the free-free distortion.

In the realistic case in which also $y_B$ should be recovered
from the CMB spectrum data, the sensitivity of the spectral distortion 
parameter recovery will result significantly affected
with respect to the case in which both 
FIRAS~II and a {\it DIMES} data are available,
with a sensitivity degradation factor of about $1.5-2$,
related to the epoch of the earlier dissipation process,
and possible significant deviations of the recovered values
from the input ones.

In addition, the rejection of ``wrong'' (i.e. different from
those assumed in the input model) models becomes less reliable
in the case of earlier dissipations at $y_h \sim {\rm some} \times 0.1$,   
being based no longer on direct $\chi^2$ analyses but on 
physical requirements about the signs of the recovered 
free-free and Comptonization distortions.

Finally, quite accurate estimates of the epochs of the
earlier processes are no longer possible.

By combining these results with those reported in Sect.~4.1
and with those presented in Burigana \& Salvaterra 2003,
we conclude that, although a significant progress
of CMB spectrum measures only at $\lambda \gsim 1$~cm
or at $\lambda \lsim 1$~cm certainly implies
a great improvement of the current knowledge
of thermal history of the Universe,
only the combination of very precise 
measures both at $\lambda \gsim 1$~cm
and at $\lambda \lsim 1$~cm will allow
to recover the spectral distortion parameters
and to map the thermal history of the Universe
under quite general conditions.

\subsection{Constraints on very high redshift processes}\label{dimes_evo}

We extend here at $z_h > z_1$ (i.e. $y_h > 5$)
the constraints on $\Delta\epsilon/\epsilon_i$ that 
would be possible to derive 
at $z_h = z_1$ ($y_h = 5$) with the combination of a {\it DIMES}-like 
and a FIRAS~II-like experiment.
We remember that at $z>z_1$ the Compton scattering 
is able to restore, after an energy 
injection, the kinetic equilibrium between matter and radiation, yielding a 
BE spectrum, and the combined effect of Compton scattering 
and photon production processes 
tends to reduce the magnitude 
of spectral distortions, possibly leading to a blackbody spectrum. 
The contribution to the thermalization process by a possible 
cyclotron emission associated to cosmic magnetic fields 
by including also stimulated emission and absorption 
(Afshordi 2002 and references therein) 
has been recently evaluated 
for realistic shapes of the distorted spectrum at early epochs
(Zizzo 2003, Zizzo \& Burigana 2003):
the strong decreasing of the frequency dependent 
chemical potential, $\mu(x)$, of the BE-like spectrum  
(Sunyaev \& Zeldovich 1970)
at the very long wavelengths relevant for the cyclotron process 
implies a large suppression of the cyclotron efficiency. 
Therefore,  
only for values of the cosmic magnetic field, $B$, 
larger than $\approx 10^{-6}$~G,
well above current observational evidencies and theoretical 
predictions, this process is found to play a significant role, 
comparable to (or larger than) the combined effect  
of radiative Compton and bremsstrahlung that
drives the thermalization process for reasonable values of $B$. 

We consider here for simplicity only the case of the simulated 
observation of a not distorted spectrum, 
that represents a good example of the possible improvements 
represented by such kinds of experiments,
directly comparable to the results based on the current data.
The comparison is shown in Fig.~2, where the results obtained
by considering only simulated data from a {\it DIMES}-like experiment 
together with the real FIRAS data 
(Burigana \& Salvaterra 2003) are also reported.

Note how the constraints on $\Delta\epsilon/\epsilon_i$ can be
improved by a factor $\sim 10 - 100$ for processes possibly occurred
in a wide range of cosmic epochs, corresponding to about a decade 
in redshift at $z$ about $10^6$.

It is also evident the improvement 
of the constraints on $\Delta\epsilon/\epsilon_i$
at $z_h \lsim {\rm few} \times 10^5$,
possible by adding FIRAS~II data to {\it DIMES} data.

Finally, note that the results based on simulated FIRAS~II and  
{\it DIMES} data are obtained 
by relaxing any assumption on $y_B$, assumed to be null 
in the other two cases (compare also the different cases reported
in Tables~1 and 2).

Of course, large energy injections are still possible at very early epochs 
close to the thermalization redshift, when primordial nucleosynthesis 
set the ultimately constraints on 
energy injections in the cosmic radiation field.

\begin{figure*}
\epsfig{figure=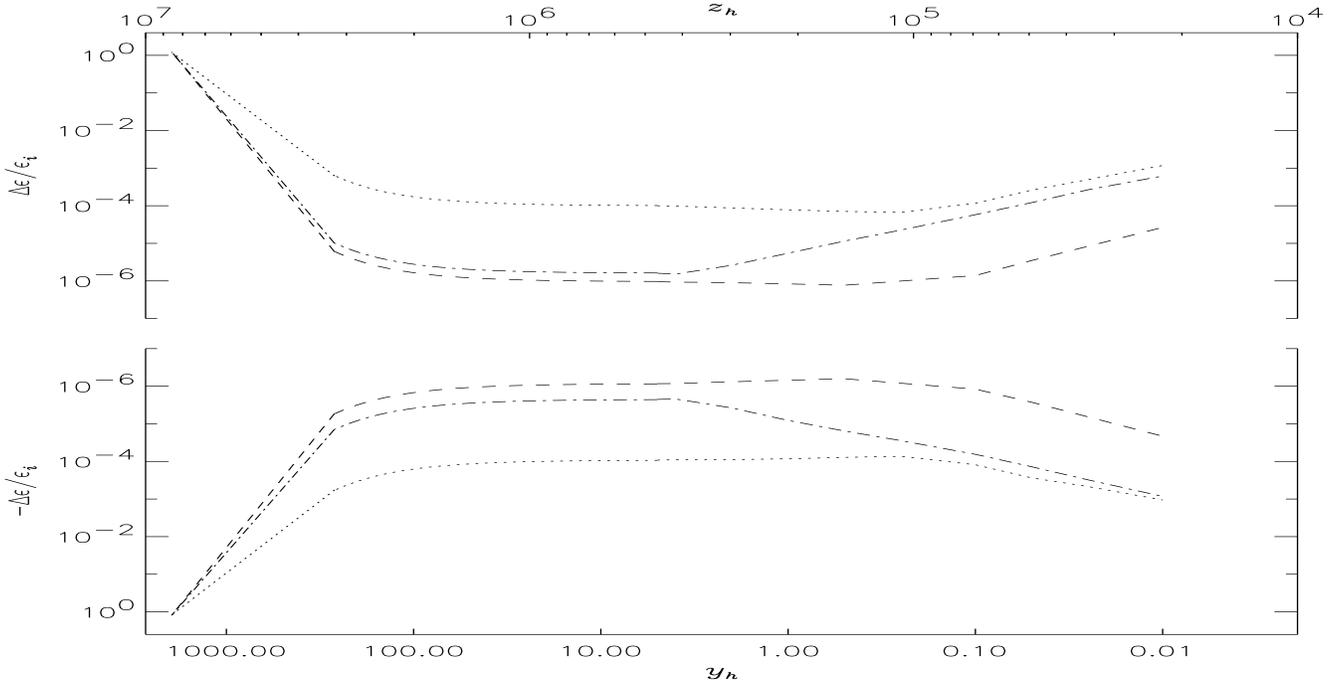,height=10cm,width=17cm}
\caption{Comparison between the constraints on the energy exchanges 
derived from current measures --
FIRAS and long wavelength data -- (dotted lines; 
in practice FIRAS data alone set the current constraints, 
see Salvaterra \& Burigana 2002), from FIRAS data jointed to 
a simulated data set from a {\it DIMES}-like experiment
(dash-dotted lines; see Burigana \& Salvaterra 2003),
and, finally, from simulated data sets from a {\it DIMES}-like experiment
jointed to a FIRAS~II-like experiment (dashed lines). 
An underlying blackbody spectrum is here assumed
for the simulated data sets.
In the first two cases (dotted lines and dash-dotted lines)
we report the constraints on  
$\Delta\epsilon/\epsilon_i (y_h)$ allowing for 
a later energy exchange at $y_h \ll 1$ but neglecting
free-free distortions (i.e. assuming $y_B =0$).
In the last case (dashed lines) we relax the
assumption $y_B =0$, i.e. we jointly consider
three kinds of spectral distortions. 
See also the text.}
\end{figure*}

\section{Implications for some classes of late astrophysical 
processes}\label{dimes_omegab}

In the previous sections the constraints on free-free and Comptonization 
distortions from the new generation of CMB spectrum experiments have been 
considered from a general point of view, 
without assuming detailed physical models for their genesis.
Such kinds of late distortions are expected in many astrophysical 
scenarios involving some energy dissipation mechanisms 
during various stages of the star and galaxy evolution 
or some particle or scalar field decay model 
(see e.g. De~Zotti \& Burigana 1992, Tegmark \& Silk 1995).
The relevance of their detection (Kogut 2003)
has been partially renewed by the {\it WMAP} satellite discovery of a 
reionization phase at relevant redshifts, mainly 
supported by {\it WMAP} detection of an excess in the 
CMB TE cross-power spectrum on large angular scale 
(multipoles $\ell \lsim 7$), indicating an optical depth to the
CMB last scattering surface of $\tau_e \simeq 0.16$ (Kogut et al. 2003). 

Without the ambition to exhaustively cover a so wide topic, 
we compare here the coupled free-free and Comptonization distortions
for some of the models recently discussed in the literature
and discuss the possibility to reveal them with the 
new generation of CMB spectrum experiments. 

Table~3 summarizes the chances of observing the Comptonization 
and free-free distortions for the four classes of processes 
discussed in the following subsections on the basis of the 
sensitivities reported in Table~2 (of course, a more optimistic
conclusion for the possibility to measure the Comptonization 
distortion, i.e. a sensitivity improvement by a factor $\simeq 2.5-3$, 
could be derived by assuming a quite strong prior 
on the earlier energy exchange; compare the last row of Table A1
with the last three rows of Table~2). 

\subsection{Free-free emitters}

Oh (1999) predicted the level of free-free distortion 
can be produced by ionized halos in the early stage of galaxy formation 
by computing the moments of the intensity distribution 
from a Press-Schechter based model. 
The predicted free-free distortion results to be
$\Delta T_{B}=3.4\times 10^{-3}$~K 
at $\simeq 2$~GHz corresponding to a free-free distortion parameter
$y_{B}\simeq 1.5\times 10^{-6}$,
well within the sensitivity of {\it DIMES}.   
The associated Comptonization distortion predicted in this work 
is below the current FIRAS upper limit by about one order of magnitude, 
and is then likely measurable by a FIRAS~II-like experiment. 
Moreover, this work assumed that free-free emitters are 
the only radio-bright sources in the sky.
The existence of radio-loud galaxies and AGNs is likely to increase the
spectral distortion due to radio sources.

\subsection{Quasar driven blastwaves}

The integrated Comptonization distortion from quasar driven
blastwaves has been recently computed by Platania et al. 2002 
through a quite general ``energetic'' approach, not particularly 
related to a detailed physical assumption. They found
$u \approx 2.4 \times 10^{-6}$, clearly measurable 
with a FIRAS~II-like experiment, with approximately 
the same fractional contribution from low-$z$ (0--2) and high-z (2--4) 
objects. The evaluation of the corresponding integrated free-free 
distortion is less model independent, being significantly sensitive 
to the density profile inside the blastwave because of the 
$n^2_e$ dependence of the free-free emission. By assuming the model 
described in Sect.~2 of Platania et al. 2002, 
we have computed the ratio between the Comptonization parameter $u$
and the free-free parameter $y_B$ integrated over the  
line-of-sight of the blastwave centre between the galaxy outer cut-off 
radius, $R_g$, approximately separating the virialized and infall regions,
and an inner radius, $r_{ff}$ ($\approx (1/10 - 1/20) \times R_g$, 
where the time required for the shock front 
to reach the outer radius of the galaxy equals the free-free cooling
timescale (it is in fact reasonable to assume that the physical conditions 
at $r<r_{ff}$ are more related to the central engine than to the 
blastwave evolution). By assuming an electron temperature 
$T_e \sim 10^6$~K ($T_e \sim 10^7$~K)
and a wide range of line-of-sight velocity 
dispersions, $\sigma \sim 100-400$~Km/s, 
we find $y_B/u \lsim {\rm few} \times 10^{-2}$ 
($\lsim {\rm few} \times 10^{-3}$) for 
objects both at $z \sim 1$ and at $z \sim 3$, while only for
low values of $T_e$ ($\sim 10^5$~K) and high values of $\sigma$
($\sim 400$) $y_B/u$ approaches the unity. A reasonable 
upper limit to $y_B$ in this model is then 
$\sim {\rm some} \times 10^{-8} - 10^{-7}$,
a value significantly smaller than the free-free distortion
predicted in the model by Oh 1999
and below or comparable to the sensitivity 
of a {\it DIMES}-like experiment.

\subsection{First stars in galaxies}

The reionization of the Universe indicated by the {\it WMAP} data
must have begun at relatively high redshift. 
Ciardi, Ferrara \& White (2003) have  studied the reionization
process using supercomputer simulations of a large and representative region of
a Universe which has cosmological parameters consistent with the {\it WMAP} 
results. The simulation follows both the radiative transfer of ionizing photons
and the formation and evolution of the galaxy population which produces them.

They have shown that the {\it WMAP} measured optical depth to electron 
scattering is easily reproduced by a model in which reionization is caused by
the first stars in galaxies with total masses of a few $\times10^9\;\Msun$. 
Moreover, the first
stars are ``normal objects'', i.e. their mass is in the range 1--50 
$\Msun$, but metal-free. Among the different model explored, the ``best'' 
{\it WMAP} value for $\tau_e$ is matched assuming 
a moderately top-heavy initial 
mass function (IMF) (Larson IMF with $M_c=5\;\Msun$) 
and an escape fraction of 20\%. A Salpeter IMF with
the same escape fraction gives $\tau_e=0.132$, which is still all the suggested
68\% confidence ranges. Decreasing $f_{esc}$ to 5\% gives $\tau_e=0.104$, 
which disagrees with {\it WMAP} only at $\simeq 1.0 - 1.5 \sigma$ level,
the exact level depending on the detailed cosmological scenario adopted to fit the WMAP 
(and auxiliary) data. 

Using the redshift evolution of the density of free electrons obtained by 
these simulations (kindly provided us by A. Ferrara), we compute the expected CMB 
spectral distortions. We obtain $u-$distortions of the order of 
$2.3\times10^{-6}$ ($4.5\times 10^{-7}$) for a constant electron temperature 
of $10^{6}$~K ($2\times 10^{4}$~K) corresponding to an energy injection of 
$\Delta\epsilon/\epsilon_i\sim 9.2\times 10^{-6}$ ($1.8\times 10^{-6}$)
well below the current FIRAS upper limits, but easily detectable by
an experiment with a sensitivity such that proposed for FIRAS II. 
We compute also the corresponding 
free-free distortion that results to be so small ($y_B\sim 10^{-11}$)
that can not be detected by any future experiment. 

In the best-fit model of Ciardi et al. (2003) reionization is essentially 
complete by $z_r\simeq 13$. This is difficult to reconcile with observations 
of the Gunn-Peterson effect in $z>6$ quasars (Becker et al. 2001; 
Fan et al. 2002). This implies a volume-average neutral fraction above 
$10^{-3}$ and a mass-averaged neutral fraction $\sim 1$\% at $z=6$. 
A fascinating (although speculative) possibility
is that the Universe was reionized twice (Cen 2003; Whyte \& Loeb 2002) with
a relatively short redshift interval in which the IGM became neutral again. 

\subsection{Particle decays}

An alternative scenario has been proposed by Hansen \& Haiman (2003). They
considered the case of a cosmologically significant particle with the 
property that it decays around a redshift $z\sim 20$. The decay products
could have energies high enough to reionize the light elements. If this 
hypothetical particle is sufficient abundant, then it may explain the high
value of the optical depth, without the need for stars to reionize the 
Universe at high redshift, but normal stellar populations could form at 
$z\simlt 10$ with the usual efficiencies and account for the completion of
reionization epoch that appears to be occurring at $z<10$.

Hansen \& Haiman (2003) proposed that this particle could be a sterile neutrino
with a mass $m_\nu\sim 200$~MeV and decay time $\tau_d \sim 4\times 10^{15}$~s.
The decaying of this neutrino can account for the electron scattering optical
depth $\tau\simeq 0.16$ (Kogut et al. 2003) measured by {\it WMAP} without
violating existing astrophysical limits on the CMB and gamma ray background.
In particular, they have shown that the expected energy injection in the CMB
of this process is $\Delta\epsilon/\epsilon_i\sim 0.4-3 \times 10^{-6}$, depending
on the fraction of the mean electron energy that goes into heating the IGM.
These values are well below the current FIRAS upper limits, but in principle
can be detected by future experiments such as FIRAS II (Fixsen \& Mather
2002). We calculate the associated free-free distortion of the CMB 
spectrum generated during a redshift interval corresponding 
to a time interval between $\sim \tau_d /10$ and $\sim 3 \tau_d$ 
for different simple thermal histories ($T_e \simeq {\rm const}$ 
or $\phi \simeq {\rm const}$) producing the above values of 
$\Delta\epsilon/\epsilon_i$. The expected $y_B$ is in the
range $0.4-1.4 \times 10^{-7}$ depending mainly on the exact amount of energy 
that goes into the IGM and only slightly on the considered thermal history of the 
process. Particularly for cases in which a large amount of the electron 
energy goes into heating the IGM, 
this distortion can be in principle observable 
by the future {\it DIMES} experiment. 

\begin{table}
\begin{center}
\begin{tabular}{lccccc}
\hline
\hline
\\
Process & Comptonization distortion & Free-free distortion \\ 
        & (mainly through FIRAS~II) & (mainly through {\it DIMES}) \\ 
\hline
FFE     & measurable          &  measurable \\
\hline
QSOB    & measurable                & $\approx$~at limit \\
\hline
FSG     & measurable                & unobservable \\
\hline
PD      & measurable                & observable  \\  
\hline
\end{tabular}
\end{center}
\caption{Chances to observe or quite accurately measure Comptonization and 
free-free distortions with the new generation of CMB spectrum space experiments
for some classes of processes.
FFE: free-free emitters; QSOB: quasar driven blastwaves;
FSG: first stars in galaxies; PD: particle decay.
Clearly, the combined effect of these (and, of course, many
others) processes should be carefully evaluated and, 
at least for the astrophysical models, 
the ``integrated'' information contained in the CMB spectrum
should be hopefully complemented by 
(or, viceversa, should complement the)  
dedicated observations (or, at least, detections) 
of the corresponding astrophysical 
sources and, possibly, foreground fluctuation analyses.
Anyway, these examples show how the combined information of 
accurate CMB spectrum measures at short and long 
wavelengths can greatly improve the reliability 
of a physical explanation of a possible observation
of a given kind of spectral distortion.}
\label{tab:proc}
\end{table}

\section{Discussion and conclusions}

We have studied the implications 
of the new generation of CMB spectrum space experiments 
for our knowledge of the thermal
history of the Universe. 

The combination of two experiments with the sensitivity and the  
frequency coverage jointly forseen for {\it DIMES} and FIRAS~II will be able to 
greatly change our vision of the capability of the CMB spectrum 
information to constrain physical processes at different cosmic ages.
The limits on the energy dissipations at the all cosmic times
accessible to CMB spectrum investigations ($z \lsim z_{therm}$)
could be improved by about two order of magnitudes and even 
dissipation processes with $\Delta\epsilon/\epsilon_i \sim 10^{-6}$
could be detected and possibly accurately studied.

We tested also the possibility to have an independent cross-check 
on the baryon density accurately measured by CMB anisotropy experiments
in presence of possible small early distortions, 
by using such accurate measures at $\lambda \lsim 1$~cm to determine the 
chemical potential and the longer wavelengths to estimate
$\Ohat_b$ through the decreasing of the equivalent thermodynamic 
temperature at increasing wavelengths, 
without finding a significant improvement with respect to
the results previously described in Burigana \& Salvaterra 2003:
only high accuracy measures at very long 
wavelengths ($\approx 50$~cm) about the minimum of the
equivalent thermodynamic temperature are able to provide 
a firm and accurate independent cross-check on $\Ohat_b$.

With a joint analysis of two dissipation processes occurring 
at different epochs, we demonstrated that 
the sensitivity and the wide frequency range jointly covered by  
{\it DIMES} and FIRAS~II would allow 
to accurately recover the two amounts of energy 
exchanged in the primeval plasma and to constrain quite well also the 
epochs of the two processes even when 
possible, a priori unknown, imprints from free-free distortions are taken into 
account in the data analysis. 
All the three distortion parameters could be in fact accurately
reconstructed in this perspective: 
the sensitivity at 95 per cent CL 
is $\simeq (5-9) \times 10^{-7}$
for the two values of $\Delta\epsilon/\epsilon_i$ and of
$\simeq 10^{-7}$ for $y_B$.

These results are possible because such levels of accuracy on a so wide 
frequency range allow to remove the approximate degeneracy both between
free-free and BE-like distortions and between 
Comptonization and BE-like distortions that remain 
in presence of future significant improvements 
only at $\lambda \gsim 1$~cm or at $\lambda \lsim 1$~cm, respectively.  
The sensitivity on 
$\Delta\epsilon/\epsilon_i$, mainly determined by 
a FIRAS~II-like experiment, improves by a factor $\simeq 1.5$ 
by adding the information from a {\it DIMES}-like experiment,
while the sensitivity on 
$y_B$, mainly determined by 
a {\it DIMES}-like experiment, improves by a factor $\simeq 1.3-2.6$, 
by adding the information from a FIRAS~II-like experiment. 

Of course, not only a very good sensitivity, but also an extreme 
control of the all systematic effects and, in particular, 
of the frequency calibration is crucial to reach these goals.

Finally, we discussed the different signatures 
imprinted on the CMB spectrum by 
some late astrophysical and particle decay models 
recently proposed in the literature and possibly related
to the reionization of the Universe 
indicated by {\it WMAP}, and 
compared them with the sensitivity of such classes of CMB space 
spectrum experiments. Different processes produce 
different levels of coupled Comptonization and 
free-free distortions and the combination of 
very accurate CMB spectrum measures at long and short 
wavelengths can in principle probe or constrain 
these models.

\section*{Acknowledgements}

We warmly thank L.~Danese, G.~De~Zotti, A.~Ferrara and P.~Platania
for numberless conversations on theoretical aspects
of CMB spectral distortions and on the ionization history. 
It is a pleasure to thank M.~Bersanelli and N.~Mandolesi 
for useful discussions on CMB spectrum observations.
Some of the calculations presented in Sect.~5 have been 
carried out on an alpha digital unix machine at the IFP/CNR in Milano
by using some NAG integration codes.

\appendix

\section{Joint analysis of two kinds of spectral distortions}

For sake of completeness we report in Table~A1 the 
constraints that can be in principle derived from a
FIRAS~II-like experiment and from the combination 
of a {\it DIMES}-like experiment with a FIRAS~II-like experiment
in the joint analysis of two kinds of spectral distortions.
These sensitivities on a couple of distortion parameters apply 
in the case in which a quite strong prior 
is assumed on the third distortion parameter
(in practice, assuming it ranging within an interval 
much smaller than the corresponding sensitivity 
quoted in Table~2).
Compare Table~A1 with Table~2.  
Although the best sensitivity on $y_B$ is reached when
the earlier energy exchange is assumed known, 
the sensitivity on $y_B$ does not depend much on the
considered case and not much degrades by considering 
two energy exchanges.
Finally, the sensitivity on one of the two energy exchanges 
is better by assuming known the other one and unknown 
the free-free distortion than by assuming known the
free-free distortion and unknown 
the other energy exchange. 

\begin{table}
\begin{center}
\begin{tabular}{lccccc}
\hline
\hline
\\
Data set & \multicolumn{4}{c}{$(\Delta\epsilon/\epsilon_i)/10^{-7}$} & $y_B/10^{-7}$ \\
\cline{2-5}
\\
 & heating at & heating at & heating at &  heating at &free-free dist. at \\
 & $y_h=5$ & $y_h=1.5$ & $y_h=0.5$ & $y_h\ll 1$ & $y_h\ll 1$ \\
\hline
\\
F-II & 1.27$\pm$8.84 & & & 0.05$\pm$4.66 & \\
F-II & & 1.00$\pm$8.02 & & 0.04$\pm$4.97 & \\
F-II & & & 0.77$\pm$6.97  & 0.04$\pm$5.45 & \\
F-II & 0.40$\pm$4.83 & & & & -2.83$\pm$8.97 \\
F-II & & 0.34$\pm$4.03 & & & -2.84$\pm$8.81 \\
F-II & & & 0.29$\pm$3.11 & & -2.84$\pm$8.57 \\
F-II & & & & 0.38$\pm$2.16 & -2.91$\pm$7.60 \\
D + F-II & 0.67$\pm$7.69 & & & 2.84$\pm$4.15 & \\
D + F-II & & 0.81$\pm$7.60 & & 0.14$\pm$4.72 & \\
D + F-II & & & 0.74$\pm$7.02 & 0.07$\pm$5.48 & \\
D + F-II & 1.13$\pm$4.00 & & & & -0.01$\pm$0.90 \\
D + F-II & & 1.00$\pm$3.38 & & & -0.04$\pm$0.88 \\
D + F-II & & & 0.81$\pm$2.67 & & -0.05$\pm$0.87 \\
D + F-II & & & & 0.60$\pm$2.08 & -0.06$\pm$0.87 \\
\\
\hline
\end{tabular}
\end{center}
\caption{Fits to two kinds of spectral distortions, jontly considered, for 
different simulated data.
D: {\it DIMES}; F-II: FIRAS II.
An underlying blackbody spectrum is here assumed.}
\label{tab:2}
\end{table}

\end{document}